\documentclass[aps,prl,twocolumn,showpacs,superscriptaddress]{revtex4-1}
\usepackage{amsmath}
\usepackage{epsfig}
\usepackage{color}
\usepackage{endnotes}
\usepackage{natbib}
\usepackage[colorlinks,breaklinks]{hyperref}
\usepackage{nicefrac}
\usepackage{bm}
\usepackage{dcolumn}
\usepackage{graphics}
\usepackage{graphicx} 
\usepackage{tikz} 
\usepackage{amsmath}
\usepackage{amssymb}
\usepackage{color}
\usepackage{changes}
\usepackage{multirow}
\usepackage{lineno}

\newcommand*{\MIT }{Massachusetts Institute of Technology, Cambridge, MA 02139, USA}
\newcommand*{\ODU}{Old Dominion University, Norfolk, VA 23529, USA}
\newcommand*{\JLAB}{Thomas Jefferson National Accelerator Facility, Newport News, VA 23606, USA}
\newcommand*{\TAU }{School of Physics and Astronomy, Tel Aviv University, Tel Aviv 69978, Israel}
\newcommand*{\NRCN}{Nuclear Research Center Negev, Be'er Sheva 84190, Israel}
\newcommand*{\GWU}{The George Washington University, Washington, DC, 20052, USA}

\newcommand*{\ANL}{Argonne National Laboratory, Argonne, Illinois 60439}
\newcommand*{\CSUDH}{California State University, Dominguez Hills, Carson, CA 90747}
\newcommand*{\CANISIUS}{Canisius College, Buffalo, NY}
\newcommand*{\CMU}{Carnegie Mellon University, Pittsburgh, Pennsylvania 15213}
\newcommand*{\CUA}{Catholic University of America, Washington, D.C. 20064}
\newcommand*{\SACLAY}{IRFU, CEA, Universit\'{e} Paris-Saclay, F-91191 Gif-sur-Yvette, France}
\newcommand*{\CNU}{Christopher Newport University, Newport News, Virginia 23606}
\newcommand*{\UCONN}{University of Connecticut, Storrs, Connecticut 06269}
\newcommand*{\DUKE}{Duke University, Durham, North Carolina 27708-0305}
\newcommand*{\DUQUESNE}{Duquesne University, 600 Forbes Avenue, Pittsburgh, PA 15282 }
\newcommand*{\FU}{Fairfield University, Fairfield CT 06824}
\newcommand*{\FERRARAU}{Universita' di Ferrara , 44121 Ferrara, Italy}
\newcommand*{\FIU}{Florida International University, Miami, Florida 33199}
\newcommand*{\FSU}{Florida State University, Tallahassee, Florida 32306}
\newcommand*{\GWUI}{The George Washington University, Washington, DC 20052}
\newcommand*{\ISU}{Idaho State University, Pocatello, Idaho 83209}
\newcommand*{\INFNFE}{INFN, Sezione di Ferrara, 44100 Ferrara, Italy}
\newcommand*{\INFNFR}{INFN, Laboratori Nazionali di Frascati, 00044 Frascati, Italy}
\newcommand*{\INFNGE}{INFN, Sezione di Genova, 16146 Genova, Italy}
\newcommand*{\INFNRO}{INFN, Sezione di Roma Tor Vergata, 00133 Rome, Italy}
\newcommand*{\INFNTUR}{INFN, Sezione di Torino, 10125 Torino, Italy}
\newcommand*{\INFNPAV}{INFN, Sezione di Pavia, 27100 Pavia, Italy}
\newcommand*{\ORSAY}{Universit'{e} Paris-Saclay, CNRS/IN2P3, IJCLab, 91405 Orsay, France}
\newcommand*{\Juelich}{Institute fur Kernphysik (Juelich), Juelich, Germany}
\newcommand*{\JMU}{James Madison University, Harrisonburg, Virginia 22807}
\newcommand*{\KNU}{Kyungpook National University, Daegu 41566, Republic of Korea}
\newcommand*{\LAMAR}{Lamar University, 4400 MLK Blvd, PO Box 10009, Beaumont, Texas 77710}
\newcommand*{\MISS}{Mississippi State University, Mississippi State, MS 39762-5167}
\newcommand*{\ITEP}{National Research Centre Kurchatov Institute - ITEP, Moscow, 117259, Russia}
\newcommand*{\UNH}{University of New Hampshire, Durham, New Hampshire 03824-3568}
\newcommand*{\NSU}{Norfolk State University, Norfolk, Virginia 23504}
\newcommand*{\OHIOU}{Ohio University, Athens, Ohio  45701}
\newcommand*{\RPI}{Rensselaer Polytechnic Institute, Troy, New York 12180-3590}
\newcommand*{\ROMAII}{Universita' di Roma Tor Vergata, 00133 Rome Italy}
\newcommand*{\MSU}{Skobeltsyn Institute of Nuclear Physics, Lomonosov Moscow State University, 119234 Moscow, Russia}
\newcommand*{\SCAROLINA}{University of South Carolina, Columbia, South Carolina 29208}
\newcommand*{\TEMPLE}{Temple University,  Philadelphia, PA 19122 }
\newcommand*{\UTFSM}{Universidad T\'{e}cnica Federico Santa Mar\'{i}a, Casilla 110-V Valpara\'{i}so, Chile}
\newcommand*{\INSUBRIA}{Universit\`{a} degli Studi dell'Insubria, 22100 Como, Italy}
\newcommand*{\BRESCIA}{Universit\`{a} degli Studi di Brescia, 25123 Brescia, Italy}
\newcommand*{\GLASGOW}{University of Glasgow, Glasgow G12 8QQ, United Kingdom}
\newcommand*{\YORK}{University of York, York YO10 5DD, United Kingdom}
\newcommand*{\VT}{Virginia Tech, Blacksburg, Virginia   24061-0435}
\newcommand*{\VIRGINIA}{University of Virginia, Charlottesville, Virginia 22901}
\newcommand*{\WM}{College of William and Mary, Williamsburg, Virginia 23187-8795}
\newcommand*{\YEREVAN}{Yerevan Physics Institute, 375036 Yerevan, Armenia}
\newcommand*{\NOWISU}{Idaho State University, Pocatello, Idaho 83209}
\newcommand*{\NOWBRESCIA}{Universit\`{a} degli Studi di Brescia, 25123 Brescia, Italy}
\newcommand*{\BEIHANG}{School of Physics, Beihang University, Beijing 100191, China}

\begin{document}


\title{$^{12}$C(e,e’pN) Measurements of Short Range Correlations in the Tensor-to-Scalar Interaction Transition Region}

\author{I.~Korover}
\affiliation{\NRCN}
\author{J.~R.~Pybus}
\affiliation{\MIT}
\author{A.~Schmidt}
\affiliation{\GWU}
\affiliation{\MIT}
\author{F.~Hauenstein}
\affiliation{\MIT}
\affiliation{\ODU}
\author{M.~Duer}
\affiliation{\TAU}
\author{O.~Hen}
\email[Contact Author \ ]{hen@mit.edu}
\affiliation{\MIT}
\author{E.~Piasetzky}
\affiliation{\TAU}
\author{L.B.~Weinstein}
\affiliation{\ODU}
\author{D.W.~Higinbotham}
\affiliation{\JLAB}
\author {S. Adhikari} 
\affiliation{\FIU}
% ...

\collaboration{The CLAS Collaboration}

\date{\today}

\begin{abstract}
  High-momentum configurations of nucleon pairs at short-distance are
  probed using measurements of the $^{12}$C$(e,e'p)$ and
  $^{12}$C$(e,e'pN)$ reactions (where $N$ is either $n$ or $p$), at
  high-$Q^2$ and $x_B>1.1$.  The data span a missing-momentum range of
  300--1000~MeV/c and are predominantly sensitive to the transition
  region of the strong nuclear interaction from a Tensor to Scalar
  interaction.  The data are well reproduced by theoretical
  calculations using the Generalized Contact Formalism with both
  chiral and phenomenological nucleon-nucleon ($NN$) interaction
  models.  This agreement suggests
  that the measured high missing-momentum protons up to $1000$ MeV/c
  predominantly belong to short-ranged correlated (SRC) pairs.  The measured $^{12}$C$(e,e'pN)$ /
  $^{12}$C$(e,e'p)$ and $^{12}$C$(e,e'pp)$ / $^{12}$C$(e,e'pn)$ cross-section ratios
  are consistent with a decrease in the fraction of proton-neutron SRC
  pairs and increase in the fraction of proton-proton SRC pairs with
  increasing missing momentum.  This confirms the transition from an
  isospin-dependent tensor $NN$ interaction at $\sim 400$ MeV/c to an
  isospin-independent scalar interaction at high-momentum around $\sim
  800$ MeV/c as predicted by theoretical calculation.
\end{abstract}

\maketitle


High momentum-transfer electron- and proton-scattering measurements,
as well as many-body \textit{ab-initio} calculations,
have shown that nucleons in the nuclear ground state temporarily form pairs with large relative momentum and smaller center-of-mass (CM) momentum.
These are called Short-Range Correlated (SRC) pairs~\cite{Atti:2015eda,Hen:2016kwk}. 
The existence and characteristics of SRC pairs are related to outstanding issues in particle, nuclear, and astrophysics, including the modification of the internal structure of nucleons bound in atomic nuclei (i.e., 
the EMC effect)~\cite{weinstein11,Hen:2016kwk,Schmookler:2019nvf,Miller:2019mae,Segarra:2019gbp},
matrix elements used to interpret searches for neutrinoless double beta
decay~\cite{Kortelainen:2007rh,Simkovic:2009pp,Benhar:2014cka,Cruz-Torres:2017sjy},
scale separation and factorization of many-body nuclear
wavefunctions~\cite{CiofidegliAtti:1995qe,Feldmeier:2011qy,Weiss:2016obx,Cruz-Torres:2019fum,Atti:2015eda},
nuclear charge radii~\cite{Miller:2018mfb}, 
and the nuclear symmetry energy governing neutron star properties~\cite{Li:2018lpy,hen15,frankfurt08b}.

A well-established feature of SRC pairs is their predominance by
proton-neutron ($pn$) pairs in the missing momentum range of
300 -- 600~MeV/c~\cite{subedi08,hen14,Duer:2018sxh,piasetzky06}.  This
results from the preference for spin-1 $pn$-pairs by
the tensor part of the $NN$ interaction, which dominates over the
scalar part of the interaction at these missing
momenta~\cite{schiavilla07, alvioli08, sargsian05}.  At higher missing
momentum calculations suggest that the scalar repulsive core should become dominant and lead to
an increased fraction of proton-proton ($pp$) SRC
pairs~\cite{Feldmeier:2011qy,Atti:2015eda,Hen:2016kwk,Ryckebusch:2018rct,Lyu:2019bxr}.
SRC measurements of this tensor-to-scalar transition provide valuable insight
into the nature of the strong nuclear interaction at short-distances.

The tensor-to-scalar transition was studied experimentally in two previous works via
measurements of the relative abundances of $pn$- and $pp$-SRC,
extracted from $(e,e'pp)$ / $(e,e'pn)$ and $(e,e'pN)$
/ $(e,e'p)$ cross-section ratios (where $N$ is either $n$ or $p$).  
Ref.~\cite{korover14} measured the
missing momentum dependence of these cross-section ratios in $^4$He out to $800$ MeV/c.
The measured $^{4}$He$(e,e'pp)$ / $^{4}$He$(e,e'pn)$ ratio was
consistent with the expected increase in the $pp$-SRC fraction with increased momenta.
However, the data has large uncertainties and the suggested increase was
due to an underlying decrease in the $^{4}$He$(e,e'pn)$ / $^{4}$He$(e,e'p)$ ratio (i.e. less $np$-SRC pairs),
whereas the $^{4}$He$(e,e'pp)$ / $^{4}$He$(e,e'p)$ ratio was overall
flat as a function of missing momentum (i.e. no increase in $pp$-SRC pairs). 

More recently Ref.~\cite{schmidt20} measured the $^{12}$C$(e,e'pp)$ /
$^{12}$C$(e,e'p)$ reaction yield ratio over the missing-momentum range
of $400$ to $1000$ GeV/c, observing a clear increase as a function of
missing momentum.  This measurement had significantly improved kinematics 
compared with Ref.~\cite{korover14}, reaching $\langle Q^2 \rangle \sim 3$--$3.5$ GeV$/c^2$ 
for large missing momentum.  However, it was limited by only measuring the
$^{12}$C$(e,e'pp)$ / $^{12}$C$(e,e'p)$ reaction yield ratio.  This
makes its interpretation subject to several theoretical assumptions
that the data itself cannot verify.  These include the assumptions
that (a) all high missing-momentum protons belonged to $2N$-SRC pairs,
and (b) reaction effects were properly accounted for, primarily
$(n,p)$ single-charge exchange (SCX) processes.  As $np$-SRC are
always more abundant than $pp$-SRCs, even a modest SCX probability can significantly
distort the $(e,e'pp)$ reaction by having a large number of observed
$(e,e'pp)$ events originating from interactions with $pn$-SRC pairs in
which the neutron undergoes SCX.  The impact of SCX on the data of
Ref.~\cite{schmidt20} can be as large as $400\%$, with a
large missing-momentum dependence (see supplementary materials
Fig. S36).  Thus, while experimentally simpler to measure, SRC studies via
the $(e,e'pp)$ reaction are subject to model-dependent assumptions and
corrections that have not yet been tested experimentally.

Here we present the results of a direct simultaneous measurement of
$pn$- and $pp$-SRC pairs using the $^{12}$C$(e,e'pN)$ and
$^{12}$C$(e,e'p)$ reactions.  The $^{12}$C$(e,e'pn)$ data are
minimally sensitive to SCX corrections due to the small fraction of
initial-state $pp$-SRC pairs, but have larger uncertainties due to the 
low neutron detection efficiency, particularly for the lower momentum 
neutrons of the lower missing momentum data.  The
$^{12}$C$(e,e'pp)$ data are more precise but sensitive to SCX
corrections.  Together, the measurement of the different reaction
channels, in combination with theoretical calculations using the
Generalized Contact Formalism
(GCF)~\cite{Weiss:2016obx,Weiss:2018tbu,Cruz-Torres:2019fum,Pybus:2020itv},
allow establishing the $2N$-SRC dominance of the measured reactions
and the observation of a scalar repulsive interaction at short
distances.

The analysis reported on herein is based on data collected in 2004 in Hall B of the
Thomas Jefferson National Accelerator Facility (Jefferson Lab) in
Virginia, USA, re-analyzed here as part of the Jefferson Lab
data-mining initiative~\cite{DataMining}.  This data comes from
measurements of 5.01 GeV electrons scattered from deuterium and carbon
targets~\cite{Hakobyan:2008zz}, detecting the scattered electrons,
knocked-out protons, and recoil neutrons in the CEBAF Large Acceptance
Spectrometer (CLAS)~\cite{Mecking:2003zu}.

CLAS utilized a toroidal magnetic field and six independent sets of
drift chambers (DCs) \cite{Mestayer:2000we}, time-of-flight (TOF)
scintillation counters~\cite{Smith:1999ii}, Cherenkov counters (CCs)
\cite{Adams:2001kk}, and electromagnetic calorimeters
(EC)~\cite{Amarian:2001zs} for charged particle detection and
identification.  Charged particle momenta were inferred from their
reconstructed trajectories within the magnetic field.  Electrons were
identified by requiring a signal in the CC, as well as a
characteristic energy deposition in the EC.  Protons were identified
through correlations between momentum and flight time.  The TOF and DC
polar angular acceptance was $8^\circ \leq \theta \leq 140^\circ$ and
the azimuthal angular acceptance ranged from 50\% at small polar
angles to 80\% at larger polar angles. The EC and CC polar angular
acceptance was limited to $<45^\circ$.

Neutrons with momenta of $200$--$1000$ MeV/c were detected in the TOF
counters by requiring a hit with energy deposition above threshold
(nominally $8$~MeV electron equivalent, or MeVee), no matching
charged-particle track (or partial track) in the drift chambers, and a TOF that
corresponded to $\beta < 0.75$.  We only considered hits reconstructed
inside a fiducial region that excluded 10~cm from the ends of all
scintillator paddles. Our results are not sensitive to the exact
exclusion region.  Neutron momenta were determined by time-of-flight, with a
typical resolution of 25--40~MeV/c.
This is the first work to measure recoil neutrons in CLAS.

The neutron detection efficiency was determined using the
over-constrained $d(e,e'p)n$ and $d(e,e'pn)$ reactions.  The
efficiency was extracted for different TOF energy deposition
thresholds in the range of $4$--$10$ MeV electron-equivalent (MeVee,
i.e., events where the neutron-induced signal measured in the TOF bar
is higher than that produced by an electron that deposited $4$--$10$
MeV in that bar), and as a function of the recoil neutron momentum
determined by the $d(e,e'p)n$ reaction.  For momenta above 400~MeV/c,
the typical efficiency was 4--5\%.  Between 200 and 400~MeV/c, the
efficiency was somewhat lower, approximately 2--3\%.  We verified that
the charged-particle veto efficiency, using the DC tracking system,
resulted in a negligible fraction of charged particles mis-identified
as neutrons due to tracking inefficiencies. 
Measured neutron yields are always shown after efficiency corrections.
See online supplementary
materials for additional details on the neutron identification,
detection efficiency, and momentum reconstruction resolution.

Similar to previous SRC studies~\cite{hen14,Duer:2018sxh,duer18,Cohen:2018gzh,schmidt20}, 
we considered events with scattered electron kinematics of four momentum transfer squared 
$Q^2 \equiv |\vec{q}|^2 - \omega^2 > 1.5$~GeV$^2/c^2$ and Bjorken scaling variable 
$x_B \equiv Q^2 / 2m_N\omega > 1.1$, where $m_N$ is the nucleon mass, while $\vec{q}$ and
$\omega$ are the 3-momentum and energy transferred to the nucleus by
the electron, respectively.  Assuming the electron scatters from a
single nucleon that does not reinteract as it leaves the nucleus with
momentum $\vec{p}_f$, the initial nucleon momentum $\vec{p}_i$ can be
approximated as equal to the measured missing-momentum: $\vec{p}_i
\approx \vec{p}_\text{miss} \equiv \vec{p}_f - \vec{q}$.  
We select $300 < p_\text{miss} <1000$~MeV/c
to enhance contributions from interactions with high initial momentum
nucleons, and require an angle between $\vec{p_f}$ and $\vec{q}$ smaller than
$25^\circ$, $0.62 < |\vec{p}_f|/|\vec{q}| < 0.96$ to select leading
nucleons.
Resonance production is suppressed by requiring that the $(e,e'p)$ reaction missing
mass, assuming scattering off a standing nucleon pair, will be smaller than the sum
of the nucleon and pion mass, i.e. $M_\text{miss} \equiv \sqrt{ (q^\mu - p_f^\mu + 2m_N)^2}
< 1.1$~GeV$/c^2$.  

If the struck nucleon is part of a $2N$-SRC pair, we interpret the
reaction through the SRC break-up model where a correlated partner
nucleon is assumed to exist as an on-shell spectator carrying momentum
$\vec{p}_\text{recoil}$.  
%
Triple coincidence $^{12}$C$(e,e'pN)$ events were selected from the
$^{12}$C$(e,e'p)$ event sample by requiring the coincidence detection of such a recoil
nucleon (proton or neutron) in CLAS with momentum
$350 < p_\text{recoil} < 1000$~MeV/c.
For neutron recoils, the neutron arrival time spectrum at the scintillators 
has a peak corresponding to the neutron events sitting on top of a similar-size 
uncorrelated random background.  
This background is uniform in hit time, allowing it to be estimated from
off-time neutrons and subtracted.  More details on the event selection
and background subtraction can be found in the online supplementary 
materials.

The $x_B > 1.1$ selection is consistent with that used in
Refs.~\cite{shneor07,subedi08,korover14,Duer:2018sxh,duer18} 
and is slightly lower than the $x_B > 1.2$ selection used by
Refs. ~\cite{hen14,Cohen:2018gzh,schmidt20}.  The lower cut value is
chosen to increase statistics; we verified that this change does not
impact our agreement with the published $(e,e'pp)$ / $(e,e'p)$ ratio of
Ref.~\cite{schmidt20} that used  $x_B > 1.2$ (supplementary materials Fig. S27).

\begin{figure} [t]
\centering
\includegraphics[width=0.495\textwidth]{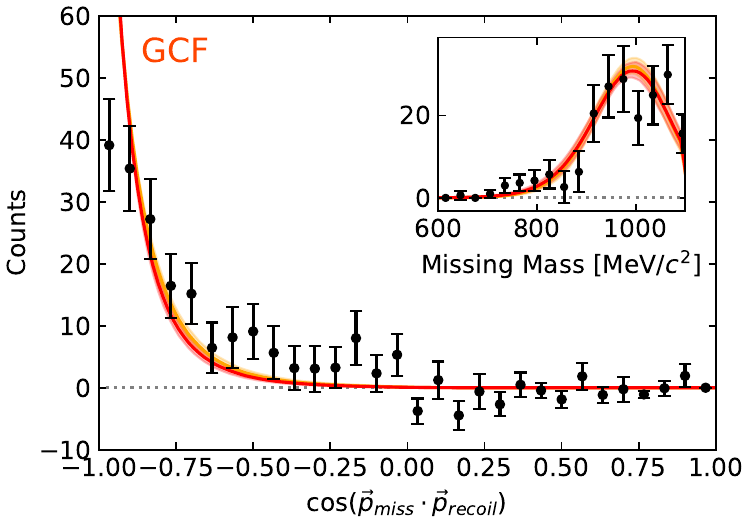}
\caption{\label{fig:OpeningAngle} Background-subtracted angular
  correlation between the reconstructed $(e,e'p)$ missing momentum
  vector ($\vec{p}_\text{miss}$) and recoil neutron momentum vector
  ($\vec{p}_\text{recoil}$), for data events passing
  $^{12}$C$(e,e'pn)$ cuts (points), compared with GCF predictions
  based on the AV18 and N2LO $NN$ interactions (darker and lighter
  bands, respectively).  Insert shows
  the background-subtracted missing mass distribution for the same
  events and calculation.  The width of the bands corresponds to the
  68\% confidence interval due to uncertainties in the model parameters. }
\end{figure}

Figure~\ref{fig:OpeningAngle} shows the cosine of the angle between
$\vec{p}_\text{miss}$ and the ``recoil'' neutron momentum $\vec{p}_\text{recoil}$ for
$^{12}$C$(e,e'pn)$ events, after random coincidence background
subtraction. While the recoil neutron selection criteria do not place
any angular requirements, the measured distribution shows a
clear back-to-back correlation characteristic of SRC breakup events.

\begin{figure*} [t]
\centering
\includegraphics[width=0.481\textwidth]{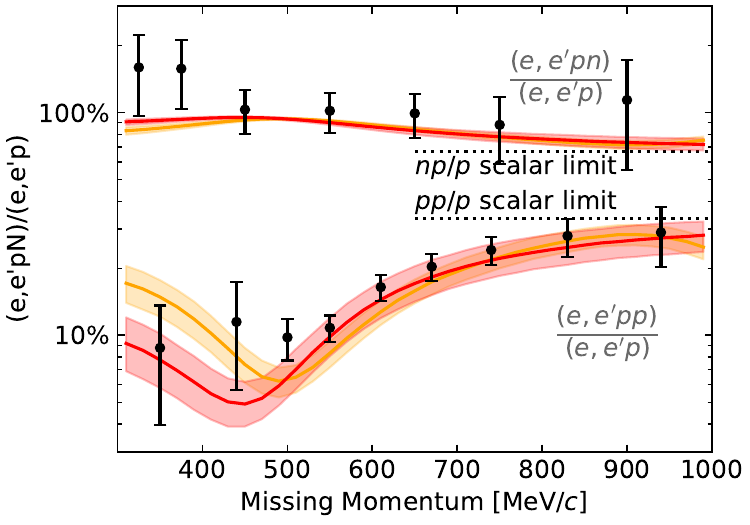}
\includegraphics[width=0.481\textwidth]{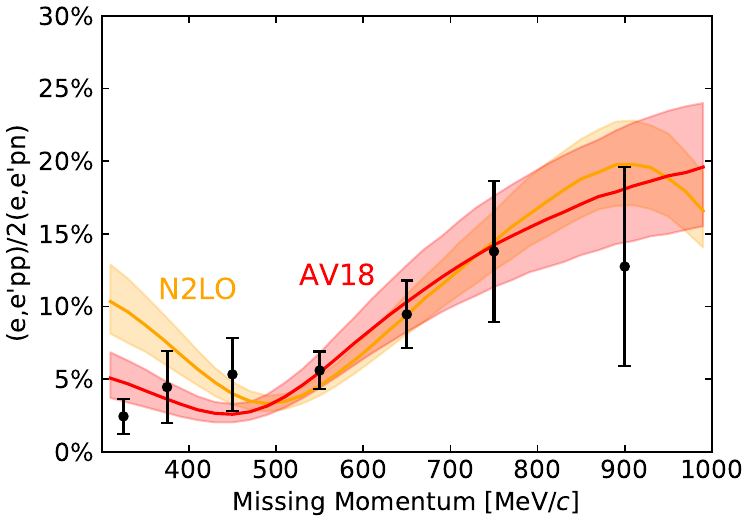}
\caption{\label{fig:ratio} Left: The measured C$(e,e'pp)/$C$(e,e'p)$
  and C$(e,e'pn)/$C$(e,e'p)$ cross-section ratios (points), compared
  with GCF predictions using
  the AV18 (darker band) and N2LO (light band) $NN$ interaction
  models.  Right: the measured C$(e,e'pp)/2\: / \: $C$(e,e'pn)$
  cross-section ratio (points), compared with GCF predictions using
  the AV18 (darker) and N2LO (lighter) $NN$ interaction models. In 
  both panels, all cross-section ratios were corrected for experimental effects
  (detector acceptance, efficiency, and resolution) as well as
  reaction effects including transparency flux lost and SCX.  The
  dashed lines marks the scalar limit obtained from a simple nucleon counting, see text for details.  The width
  of the GCF calculation bands shows their 68\% confidence interval
  due to uncertainties on the model parameters.  The data error bars
  show the quadratic sum of the statistical uncertainty and
  systematic uncertainty associated with the correction of experimental 
  effects (see online supplementary materials for details on the uncertainty 
  estimation). Not shown are the normalization uncertainties on the data; the C$(e,e'pp)/$C$(e,e'p)$
  and C$(e,e'pn)/$C$(e,e'p)$ data include a 5\% uncertainty resulting from 
  transparency corrections, and the C$(e,e'pn)$ data include correlated 
  systematic uncertainties as listed in Table S3.}
\end{figure*}

The measured distributions show good agreement with theoretical predictions
based on the GCF~\cite{Weiss:2016obx,Weiss:2018tbu,Cruz-Torres:2019fum,Pybus:2020itv}
using both the AV18~\cite{wiringa95} and N2LO(1.0)~\cite{Gezerlis:2013ipa} $NN$ interaction models.  

The GCF assumes scale-separation between the short-distance interactions
within an SRC pair, and the long-range interactions between the pair
and the rest of the nucleus, as well as their mutual separation from
the ultra-short distance scale associated with the high-energy virtual
photon probe. 
With this in mind, Ref.~\cite{Weiss:2018tbu} suggested a factorized approximation for the correlated continuum region of the nuclear spectral function, that can be used in factorized models of the scattering cross-section at large momentum transfer kinematic~\cite{kelly96}. Here the hard break-up of an SRC pair is assumed to proceeds via a reaction in which the virtual photon is absorbed by a single nucleon in an SRC pair, knocking it out of the nucleus and leaving its correlated partner nucleon to recoil from the nucleus~\cite{Weiss:2018tbu,Pybus:2020itv}.

For completeness we note that beyond the use of the specific GCF model for the spectral function, the reaction model used herein adopts a high-resolution theoretical description of high-momentum transfer reactions where the reaction is modeled using one-body operators and correlation effects are embedded in the nuclear wave function. While constituting a valid simple reaction picture that is consistent with both data and various ab initio calculations, it is not the only possible description of our data. Unitary freedom allows shifting the explicit effects of two-body correlations from nuclear wave functions to the interaction operators while keeping the calculated cross-section invariant~\cite{More:2017syr}. Thus, theoretical studies can also use our data to study complementary factorized models~\cite{Ryckebusch:2018rct} and/or constrain many-body reaction operators used in low-resolution nuclear theory calculations.

Several ingredients are necessary to construct the GCF based factorized
cross-section~\cite{Pybus:2020itv}. We used the off-shell
electron-nucleon cross-section from Ref.~\cite{DeForest:1983ahx}.  
Nuclear contacts~\cite{Weiss:2016obx,Weiss:2018tbu,Cruz-Torres:2019fum}, 
and the possible excitation range of the residual $A-2$ nuclear system $E^*$
are the same as in Ref.~\cite{schmidt20}.  
The pair CM momentum distribution is assumed to be a three-dimensional Gaussian~\cite{CiofidegliAtti:1995qe,Colle:2013nna} with a characteristic width taken from Ref.~\cite{Cohen:2018gzh}.
Additionally, we
accounted for Final State Interactions (FSIs) including Single Charge
Exchange (SCX) and nuclear transparency using the Glauber
approximation from Ref.~\cite{Colle:2015lyl}.  The transparency
correction is a simple overall scale factor and was previously shown
to well-reproduce experimental
data~\cite{hen12a,colle15,Duer:2018sjb}.  However, the SCX corrections
affect the missing-momentum dependence of the data, are less certain,
and were not validated experimentally. Therefore, obtaining a
consistent picture from analysis of both $(e,e'pn)$ and $(e,e'pp)$
data with minimal and maximal SCX sensitivity, respectively, is
crucial for a reliable interpretation of experimental data.

Systematic model uncertainties associated with the GCF predictions were
estimated by repeating the theoretical calculations with randomly
sampled model parameters from a distribution centered around the
parameter's nominal value with a width defined by its uncertainty. We
also considered two different prescriptions for the off-shell electron-nucleon cross-section 
known as cc1 and cc2 from Ref.~\cite{DeForest:1983ahx}.

Supplementary materials Fig.~S35 shows comparisons between the GCF calculations 
and the measured $p_\text{miss}$-dependence of the $^{12}$C$(e,e'pn)$ /
$^{12}$C$(e,e'p)$, $^{12}$C$(e,e'pp)$ / $^{12}$C$(e,e'p)$ and
$^{12}$C$(e,e'pp)$ / $^{12}$C$(e,e'pn)$ yield ratios. The data
are corrected for nuclear tranparency. Since this correction has no
$p_\text{miss}$-dependence, it only changes the overall scale. 
The data and calculations are in good agreement.

To extract cross-section ratios from the ratios of measured event yields,
we corrected for SCX effects, nuclear transparency, experimental acceptance,
and the efficiency of the event selection criteria. These corrections were determined 
by comparing the GCF cross-section to a detailed Monte Carlo simulation that used
the GCF cross-section as input. Simulated events were propagated through a model
of the CLAS detector that included acceptance, efficienty, and resolution effects,
and were then required to pass the exact same event selection criteria.
The detector and detector+SCX correction factors are shown in supplementary materials Fig. S36.
Further details can be found in Ref.~\cite{Pybus:2020itv}.

The uncertainty on the acceptance
correction combined the systematic uncertainty of the GCF model,
described above, with uncertainty on the acceptance coming from limited
knowledge of the spectrometer momentum resolution. This was treated
by varying the detector model's momentum resolutions for electrons,
protons, and neutrons within uncertainties in the same manner as the 
GCF model parameters.


Figure~\ref{fig:ratio} shows the resulting $^{12}$C$(e,e'pn)$ /
$^{12}$C$(e,e'p)$, $^{12}$C$(e,e'pp)$ / $^{12}$C$(e,e'p)$, and
 $^{12}$C$(e,e'pp)/2\; / \;  ^{12}$C$(e,e'pn)$ cross-section ratios as a
function of $p_\text{miss}$.  The data are compared with GCF
calculations.  For $p_\text{miss} > 400$ MeV/c the calculations agree
well with the measured data for either $NN$ potential.  This agreement
supports the validity of the GCF description of the nuclear ground state at high-momentum.
For $300 < p_\text{miss} < 400$ MeV/c, especially for the  $^{12}$C$(e,e'pp)/2\; / \;  ^{12}$C$(e,e'pn)$
ratio, the AV18 calculation agrees well with the data but the N2LO
calculation does not.  This missing-momentum region is most sensitive
to the details of the dip in the $pp$ wave function which is absent
for spin-1 $pn$ pairs due to the tensor
interaction~\cite{schiavilla07, alvioli08, sargsian05}.  This dip has
slightly different characteristics for AV18 and N2LO,
possibly owing to the N2LO interaction's short-distance regulator~\cite{Cruz-Torres:2019fum}.
On the other hand the GCF is an asymptotic model and the observed discrepancy appears near 
the lower edge of its applicability~\cite{Weiss:2016obx,Cruz-Torres:2019fum}
and should thus be studied in greater detail by future works.
At the highest missing momenta the data agree with the scalar limit prediction 
where the number of spin-$1$ $pn$ SRC pairs should be three times the number of 
spin-$0$ $pp$, $pn$ and $nn$ pairs, owing to the three possible spin orientations.

\begin{figure} [t]
\centering
\includegraphics[width=0.495\textwidth]{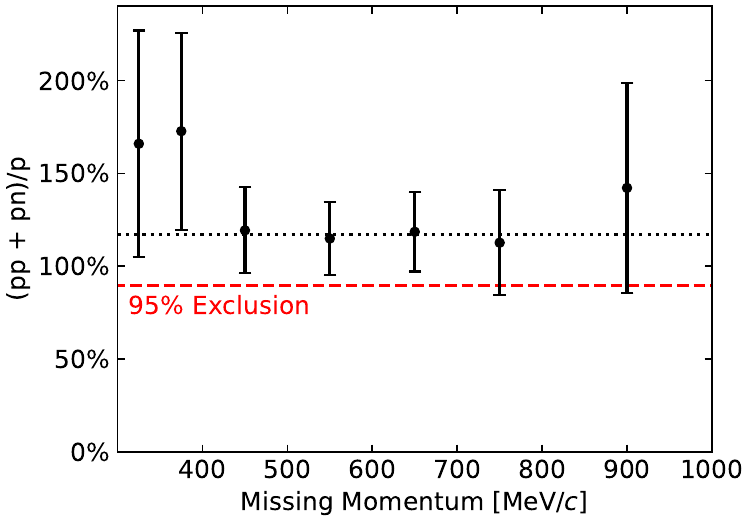}
\caption{\label{fig:SRC_Fraction} The fraction of high-momentum
  protons with a measured recoil partner nucleon for $^{12}$C: the
  measured ratio of $(e,e'pp)+(e,e'pn)$ events to $(e,e'p)$ events as
  a function of $\vec{p}_\text{miss}$.  The dotted (black) line is the
  best constant fit $117\%$. The dashed (red) line shows the
  $95\%$CL lower bound on the 2N-SRC contribution to the
  $^{12}$C$(e,e'p)$ strength in the measured missing-momentum range
  at 89\%.
}
\end{figure}

Last, Fig.~\ref{fig:SRC_Fraction} shows the fraction of $(e,e'p)$
events with a correlated recoil nucleon, i.e., the $[^{12}$C$(e,e'pp)$ +
$^{12}$C$(e,e'pn)]$ / $^{12}$C$(e,e'p)$ cross-section ratio.  Unlike
the individual $^{12}$C$(e,e'pN)$ / $^{12}$C$(e,e'p)$ ratios, this sum
is insensitive to SCX corrections.  The data show no significant missing-momentum
dependence and imply that within uncertainties, all high-initial-momentum protons are
accompanied by a correlated spectator recoil nucleon and therefore belong
to a 2N-SRC. The mean of the data points exceeds 100\%.
This is consistent given the large correlated normalization uncertainty of approximately 12\% 
that is driven by uncertainties in the neutron detection efficiency and transparency correction
that will both take all data points up or down and thus can lead to an 'unphysical' mean value for the data (see supplementary materials for details). 
At the 95\% confidence level the data exclude contributions 
from sources other than 2N-SRCs above 11\%. This bound is determined while accounting for both data statistical and systematic uncertainties, with the latter including both point-to-point and correlated normalization uncertainties.

To conclude, we report on new measurements of the $^{12}$C$(e,e'pn)$
reaction, and improved measurements of the $^{12}$C$(e,e'pp)$ and
$^{12}$C$(e,e'p)$ reactions at very high missing momentum.  The data
are used to study the evolution of the isospin dependence of $NN$-SRCs
via the C$(e,e'pp)/2\: / \: $C$(e,e'pn)$ cross-section ratio,
and the dominance of high-momentum nucleons by 2N-SRC pairs via the
$(e,e'pn)$ / $(e,e'p)$ and $(e,e'pp)$ / $(e,e'p)$ cross-section ratios and their sum.  
The data are compared with GCF calculations using the AV18 and N2LO
interactions.  The data agrees with both calculations for
$p_\text{miss} > 400$ MeV$/c$, but disagrees with the N2LO-based
calculation for $300 < p_\text{miss} < 400$ MeV$/c$, 
near the lower edge of the applicability of the GCF~\cite{Weiss:2016obx,Cruz-Torres:2019fum}.

The overall good agreement of the GCF calculation with both $^{12}$C$(e,e'pn)$ and
$^{12}$C$(e,e'pp)$ data indicates that, within the uncertainty of the
data, the measured reactions are dominated by interactions with
$NN$-SRC pairs and that reaction effects such as SCX, which has a
large impact on the $^{12}$C$(e,e'pp)$ channel but a small impact on
the $^{12}$C$(e,e'pn)$ channel, are sufficiently well modeled.

The combination of all data and calculations confirms the observation
of a transition of the $NN$ interaction from a tensor-dominated region
around relative momenta of $400$ MeV/c to a predominantly scalar
interaction around $800$ MeV/c, validating the use of the $NN$
interactions examined here at high-momentum / short distance regimes. 
Future extensions of the GCF to three-nucleon
correlations as well as forthcoming measurements~\cite{clas12src} of
three-nucleon knockout reactions $A(e,e'pNN)$ will allow similar
studies of the short-distance three-body interactions that are needed
for a complete description of neutron stars~\cite{Gandolfi:2011xu}.

\begin{acknowledgements}
We acknowledge the efforts of the staff of the Accelerator and Physics Divisions at Jefferson Lab
that made this experiment possible. 
The analysis presented here was carried out as part of the Jefferson Lab Hall B Data-Mining
project supported by the U.S. Department of Energy (DOE). The research was also supported also by
the National Science Foundation, the Pazy Foundation, the Israel Science Foundation
and the United Kingdom's Science and Technology Facilities Council (STFC). The 
Jefferson Science Associates operates the Thomas Jefferson National Accelerator Facility for the
DOE, Office of Science, Office of Nuclear Physics under Contract No. DE-AC05-06OR23177.
\end{acknowledgements}

\bibliography{../../../../references.bib}

\end{document}